\documentclass[]{iopart}

\usepackage{caption2}
\usepackage{graphicx}
\renewcommand{\captionlabelfont}{\bf}
\renewcommand{\captionfont}{\small}
\newcommand{\be}{\begin{equation}}
\newcommand{\ee}{\end{equation}}
\newcommand{\ber}{\begin{eqnarray}}
\newcommand{\eer}{\end{eqnarray}}
\newcommand{\de}{\end{equation*}}
\newcommand{\cer}{\begin{eqnarray*}}
\newcommand{\der}{\end{eqnarray*}}

\begin{document}
\title[Generation and evolution of quantum vortex states]{Entanglement by linear SU(2) transformations: generation and evolution of quantum vortex states}
\author{G S Agarwal\footnote{On leave from Physical Research Laboratory, Navrangpura, Ahmedabad 380 009,
India}$^1$ and J Banerji$^2$}
\address {$^1$ Department of Physics, Oklahoma State University,
Stillwater, OK-74078, USA}
\address {$2$ Quantum Optics and Quantum Information Group, Physical Research Laboratory, Navrangpura, Ahmedabad 380 009,
India} \ead{jay@prl.res.in}

\begin{abstract}
We consider the evolution of a two-mode system of bosons under the
action of a Hamiltonian that generates linear SU(2)
transformations. The Hamiltonian is generic in that it represents
a host of entanglement mechanisms, which can thus be treated in a
unified way. We start by solving the quantum dynamics analytically
when the system is initially in a Fock state. We show how the two
modes get entangled by evolution to produce a coherent
superposition of vortex states in general, and a single vortex
state under certain conditions. The degree of entanglement between
the modes is measured by finding the explicit analytical
dependence of the Von Neumann entropy on the system parameters.
The reduced state of each mode is analyzed by means of its
correlation function and spatial coherence function. Remarkably,
our analysis is shown to be equally as valid for a variety of
initial states that can be prepared from a two-mode Fock state via
a unitary transformation and for which the results can be obtained
by mere inspection of the corresponding results for an initial
Fock state. As an example, we consider a quantum vortex as the
initial state and also find conditions for its revival and charge
conjugation. While studying the evolution of the initial vortex
state, we have encountered and explained an interesting situation
in which the entropy of the system does not evolve whereas its
wave function does. Although the modal concept has been used
throughout the paper, it is important to note that the theory is
equally applicable for a two-particle system in which each
particle is represented by its bosonic creation and annihilation
operators.\\\\

(Figures in this article are in colour only in the electronic
version)
\end{abstract}
\section{Introduction\label{1}}
 Nonclassical properties of quantum states are actively being
studied for their relevance in quantum computation. It is known
that quantum entanglement is the key to performing communication
and information processing tasks that cannot be realized
classically. For this reason, there has been a surge of activity
towards preparing, identifying and quantifying entangled
systems\cite{review}.

An important source of quantum entanglement has been the
polarization-entangled two-photon states generated from type-II
phase-matched parametric down conversion \cite{pdc}. A variety of
other entangled states can be produced by using various polarizing
components. More recently, the subject of quantum information
processing has been given a new direction with the realization
that a number of quantum logic operations can be performed using
single photons and methods of linear optics \cite{milburn}. Even a
method for quantum teleportation was proposed and implemented
\cite{demartini}. Clearly one needs to examine, in full
generality, the question of transformation of an arbitrary input
state by a device which can mix different states.

We note that a number of special cases for the generation of
entanglement using linear optical devices have been investigated.
Huang and Agarwal \cite{huang} considered multimode systems
described by a Hamiltonian that is quadratic in the mode
operators. They derived conditions for the generation of an
entangled state when the input state was represented by a Gaussian
density matrix. Their treatment covered a large class of states
including squeezed coherent states and even states with thermal
noise. However they did not consider the case of input fields in
Fock states. More recently and more specifically, Kim et al
\cite{knight} examined the question of the generation of entangled
states by a beam splitter using Fock states as input fields.

In this paper we specialise to intensity- or number-preserving
linear transformations belonging to the SU(2) group. For two-mode
states characterized by the annihilation operators $a$ and $b$,
such transformations can be generated by evolution under a
Hamiltonian of the form
\begin{equation} H=g(a^\dagger
b e^{i\phi} +h.c)+\Omega (a^\dagger a -b^\dagger b)\label{eq1}
\end{equation}
where $g$ and $\Omega$ are real constants. Introducing the
generators of the SU(2) group as \be \label{eq62}
 \fl J_1 = (a^\dagger b + ab^\dagger)/2,\qquad
 J_2 = (a^\dagger b - ab^\dagger)/2i,\qquad
 J_3 =  (a^\dagger a - b^\dagger b)/2,
 \ee
 the Hamiltonian (\ref{eq1}) can be rewritten in the form
 \begin{equation}\label{eq63}
 H=v_1 J_1+v_2J_2+v_3J_3\end{equation}
 with
 \begin{equation}\label{eq64}
 v_1=2 g \cos\phi,\qquad v_2=-2g\sin\phi,\qquad v_3=2\Omega.\end{equation}

Motivation for the present work comes from the realization that a
Hamiltonian of the form (\ref{eq1}) can represent a host of
entanglement mechanisms which can thus be treated in a unified
way. Several examples are given as follows.

The beam splitter, used by many authors as an entangler \cite{bs}
can be described by (\ref{eq1}) for $\Omega=0$ if one defines its
amplitude reflection and transmission coefficients by $\cos g$ and
$\sin g$ respectively while $\phi$ denotes the phase difference
between the reflected and transmitted fields .

The parametric frequency conversion by a strong pump field (of
frequency $\omega$) in $\chi^{(2)}$ material can also be represented
by an interaction Hamiltonian of the form (\ref{eq1})
\cite{louisell} with $\Omega=0$. Here $ a$ and $b$ are the
annihilation operators for the signal (of frequency $\omega_a$) and
the idler (of frequency $\omega_b$) respectively, $g$ is a coupling
constant that depends on the amplitude of the pump mode and
$\phi=\Delta\omega t$ where $\Delta\omega=\omega
+\omega_b-\omega_a$. We should note, however, that this Hamiltonian
does not support parametric down conversion.

Polarizing elements such as half- and quarter wave plates also can
act as entangling devices. Quantum mechanically, polarized light
is represented by a pair of orthogonal polarization modes
(described by boson mode operators $a$, $b$), or as points on the
Poincar\'{e} sphere. The effect of a polarizing element on the
field is a SU(2) transformation of the mode operators which
corresponds to rotations on the Poincar\'{e} sphere. The
transformations are generated by Hamiltonians of the form
(\ref{eq1}).

Finally, following the work of Wineland et al \cite{wineland}, we
consider a single laser cooled ion confined in a two dimensional
harmonic trap. The internal and motional degrees of freedom of the
ion can be coupled by applying two classical laser beams. If $a$
and $b$ represent the two oscillatory modes of the ion's quantized
motion and $\phi$ denotes the difference in phase between the two
applied fields, then, under certain conditions \cite{brif} the
Hamiltonian for the ion's motion will be of the form (\ref{eq1})
in the interaction picture.

The present work is also relevant in the context of parallel
developments in the field of optical vortices. An optical vortex
of order $l$ centered at the origin ($r=0$) has a field
distribution of the form $F(r) \exp(il\phi)$. The distribution is
such that the field intensity tends to zero as $r\to 0$ whereas
the phase shift in one cycle around the origin is $2\pi l$ where
$l$ is an integer. The azimuthal mode index $l$ has a physical
meaning in that the vortex carries  an orbital angular momentum of
$l \hbar$ per photon \cite{pra92_8185}. This angular momentum can
be imparted to microscopic particles in order to manipulate them
optically \cite{OL96_827,prl95_826}. In recent years, this
understanding has led to considerable interest in the generation
and study of optical vortices  both in free space
\cite{allenreview} and in guided media\cite{science,jayvortex}.

A physically realizable field distribution that contains optical
vortices is a higher-order Laguerre-Gaussian ($LG$) beam whose
waist-plane field amplitude is given by
\cite{oc93_123}\begin{equation} \fl u^{LG}_{mn}(x,y,\omega)
=\sqrt{{2\over \pi \omega^2}} {(-1)^p p! \over \sqrt{m!n!}}
e^{-i\theta(m-n)} (r\sqrt{2}/\omega)^{|m-n|}L_p^{|m-n|}
(2r^2/\omega^2) e^{-r^2/\omega^2}\label{eq0} \end{equation} where
$r^2=x^2+y^2$, $\theta=\arctan (y/x)$, $\omega$ is the beam waist,
$p = \min (n,m)$ and $L_p^l(x)$ is a generalized Laguerre
polynomial. $LG$ beams can be produced directly from a laser
\cite{oc94_161}. In fact, in a hydrodynamic formulation of laser
beam dynamics in terms of LG modes, vortices were found to occur in
transverse laser patterns \cite{brambilla1,brambilla2}. Usually
however, $LG$ beams are produced by the conversion or combination of
Hermite-Gaussian ($HG$) beams that are emitted by most laser
cavities. This is made possible because of the fact that any LG mode
can be expressed in terms of HG modes. The waist-plane amplitude of
the HG modes has the form
\begin{equation}\label{three} u^{HG}_{n,m}(x,y,
\omega)=\Phi_{n}(x,\omega)\Phi_{m}(y,\omega)\end{equation} where
\begin{equation}\label{four} \Phi_{n}(x,\omega)= \left({\sqrt{2}\over
\sqrt{\pi} 2^n w n!}\right)^{1/2} H_n(\sqrt{2} x/w) \exp
(-x^2/w^2) \end{equation} and $H_n(x)$ is a Hermite polynomial.
The decomposition of a LG mode in terms of HG modes is given as
\cite{allenreview} \numparts
\begin{eqnarray}
\label{six} u^{LG}_{n,m}(x,y,\omega) & = & \sum_{k=0}^{m+n} i^k
b(n,m,k) u^{HG}_{m+n-k,k}(x,y,\omega)\\
b(n,m,k) & = & \sqrt{{(n+m)! k!\over 2^{n+m} n! m!}}{1\over k!}
{d^k\over dt^k}[(1-t)^n (1+t)^m]\rfloor_{t=0}
\end{eqnarray}
\endnumparts

The vortices as discussed above appear on the transverse amplitude
profile of {\it classical} wave fields. Vortices can also occur in
the configuration space representation of quantum systems of
matter or radiation. Since the HG modes are also the energy
eigenfunctions of a quantum oscillator, quantum vortices should
arise in the study of wave packets of a quantum system that could
be a two-dimensional harmonic oscillator like an ion in a
two-dimensional trap. For a two-mode radiation field characterized
by the annihilation operators $a$, $b$,  and represented by a
state vector $\vert\psi\rangle$, the quantum vortex will appear in
the quadrature distribution $\vert\langle
x,y\vert\psi\rangle\vert^2$ where $\vert x,y\rangle$ is the
eigenvector of $(a+a^\dagger)/\sqrt{2}$ and
$(b+b^\dagger)/\sqrt{2}$. Quadrature distributions can be measured
by a homodyne method\cite{leonhardt}. Vortices of matter will
appear in the configuration space probability distribution.
Recently it has been shown that the HG and LG modes are unitarily
related\cite{simon} and the Poincare sphere\cite{poincare_cl}
representing LG beams has an underlying SU(2)
structure\cite{poincare_qm}. The Hamiltonian (1) is therefore
ideally suited to explore the possibility of generating quantum
vortices.

The objective and the plan of the paper are as follows. In section
\ref{2} we obtain the state vector and the wave function of a
two-mode system which is initially in a Fock state and is acted
upon by the Hamiltonian (\ref{eq1}). We show how the two modes get
entangled by evolution and under certain conditions evolve into a
vortex state. The degree of entanglement between the modes is
measured by finding the dependence of the von Neumann entropy on
the system parameters. In section \ref{3} the above analysis is
carried out when the two-mode system is initially in a state that
can be obtained from a Fock state via a unitary transformation. As
an example, a quantum vortex is used as the initial state. We also
find conditions for the revival and the charge conjugation of the
vortex. In section \ref{4} we consider the structure of the
reduced state of each mode. The paper ends with concluding remarks
in section \ref{5}.
\section{Generation of quantum entanglement and creation of a quantum vortex using an initial two-mode Fock state\label{2}}

\subsection{Evolution of the state vector\label{2.1}}
Let us consider the evolution of a two-mode Fock state $\vert
N-j,j\rangle$ when the Hamiltonian is given by (\ref{eq1}) and the
total number ($N$) of photons in the two modes is constant. The
resulting state $|\psi_{Nj}(t)\rangle=U(t)\vert N-j,j\rangle$ can
be obtained by the use of the disentangling theorem. In what
follows, we use a different method. We write $|N-j,j\rangle$ as
\begin{equation}\label{eq2} |N-j,j\rangle={(\hat{a}^\dagger)^{N-j}
(\hat{b}^\dagger)^{j}\over \sqrt{N-j! j!}}|0,0\rangle
\end{equation} and define a new pair of operators
\begin{equation}\label{eq3} {\hat{a}(t)\choose \hat{b}(t)}
=U^\dagger (t) {\hat{a}\choose \hat{b}} U(t) \end{equation} where
$U(t)=\exp(-iHt)$ is the time evolution operator. Then
$|\psi_{Nj}(t)\rangle$ can be written in a compact form as
\begin{equation}\label{eq4}
|\psi_{Nj}(t)\rangle={[\hat{a}^\dagger(-t)]^{N-j}
[\hat{b}^\dagger(-t)]^{j}\over \sqrt{N-j! j!}}|0,0\rangle
\end{equation} Note that the state at time $t$ is obtained by using the operators evaluated at time $-t$. The explicit expressions for $\hat{a}(t)$ and
$\hat{b}(t)$ can be obtained by solving the Heisenberg equations
for the operators. We get \begin{equation}\label{eq5}
{\hat{a}(t)\choose \hat{b}(t)}= {\bf
V}{\hat{a}(0)\choose\hat{b}(0)} ={\bf V}{\hat{a}\choose\hat{b}}
\end{equation} where ${\bf V}=\{v_{ij}\}$ is a $2\times 2$ unitary matrix.
Setting $\sigma=\sqrt{\Omega^2+g^2}$ and $\Omega=\sigma
\cos\Theta$, the matrix elements are written as
\begin{eqnarray}\label{eq6}
v_{11} = \cos \sigma t - i \cos\Theta \sin \sigma t, &&
\qquad v_{12} = - ie^{i\phi}\sin\Theta\sin \sigma t,\nonumber\\
v_{21} =  - ie^{-i\phi}\sin\Theta\sin \sigma t, &&\qquad  v_{22} =
\cos \sigma t + i \cos\Theta\sin \sigma t. \end{eqnarray} Note
that
\begin{equation}\label{eq7} v_{11}=v^*_{22},\quad
v_{12}=-v^*_{21}\quad\hbox{and}\quad \vert v_{21}\vert^2+\vert
v_{22}\vert^2=1. \end{equation} Substitution in (\ref{eq4})
followed by binomial expansion and the use of (\ref{eq2}) yields
\begin{equation}\label{eq8} |\psi_{Nj}(t)\rangle=\sum_{m=0}^{N-j}\sum_{n=0}^{j}
b_{mn}|N-(m+n), m+n\rangle \end{equation} where
\begin{equation}\label{eq9} \fl b_{mn}={N-j\choose m} {j\choose
n} {N\choose N-j}^{1/2} {N\choose
m+n}^{-1/2}(v_{11})^{N-j-m}(v_{21})^m (v_{12})^{j-n} (v_{22})^n.
\end{equation} {\it The two modes in the state $|\psi_{Nj}\rangle$ are
entangled in the sense that the above double sum cannot be reduced
to the product of two single-mode summations}.

It is instructive to briefly mention the case when the two modes are
initially in a Glauber coherent state $\vert \alpha, \beta\rangle$.
Since the Hamiltonian (\ref{eq1}) conserves photon numbers, the
state at time $t$ will also be a coherent state:
\begin{equation}\label{eq10} U(t)\vert \alpha, \beta\rangle=\vert \alpha (t),\beta
(t)\rangle. \end{equation} Applying (\ref{eq5}) on $\vert \alpha,
\beta\rangle$, we immediately obtain
\begin{equation}\label{eq11} {\alpha(t)\choose \beta (t)}= {\bf
V}{\alpha\choose \beta}.\end{equation} Furthermore, the unitarity
of ${\bf V}$ ensures that
\begin{equation}\label{eq12}\vert\alpha(t)\vert^2+\vert\beta(t)\vert^2=
\vert\alpha\vert^2+\vert\beta\vert^2.\end{equation} Thus {\it no
entanglement occurs if each input mode is in a coherent state}.

In what follows, we exploit coherent states as generating
functions of number states to reduce the double sum in (\ref{eq8})
to a single sum. Expanding both sides of (\ref{eq10}) in number
states and recalling that $\vert \alpha(t)\vert^2 +\vert
\beta(t)\vert^2=\vert \alpha\vert^2 +\vert \beta\vert^2$, we get
the relation \begin{equation}\label{eq13} \sum_m\sum_n {\alpha^m
\beta^n\over \sqrt{m!n!}} U(t)\vert m,n\rangle=\sum_p\sum_q
{\alpha^{p+q}\over \sqrt{p!q!}} \xi_{pq}(\tau)\vert p,q\rangle
\end{equation} where $\tau=\beta/\alpha$ and \be\label{eq14}
\xi_{pq}(\tau) = (v_{11}+v_{12}\tau)^p (v_{21}+v_{22}\tau)^q =
\sum_{k=0}^{p+q} {\tau^k \over k!}
\partial_\tau^{(k)} \xi_{pq}(\tau)\rfloor_{\tau\to 0}.\ee Substituting
in (\ref{eq13}) and equating the coefficient of $\alpha^{N-j}
\beta^j$, one gets \begin{equation}\label{eq15}
|\psi_{Nj}(t)\rangle=U(t)\vert N-j,j\rangle=\sum_{q=0}^N
C^{(q)}_{Nj} \vert N-q,q\rangle \end{equation} where
\begin{equation}\label{eq16} C^{(q)}_{Nj}={1\over j!}\left[{(N-j)!
j! \over (N-q)!q!}\right]^{1/2}
\partial_\tau^{(j)} \xi_{N-q,q}(\tau)\rfloor_{\tau\to 0}.\end{equation}
Some useful properties of $\vert C^{(q)}_{Nj}\vert^2$ are derived
in appendix A.
\subsection{ The wave function-- a coherent superposition of vortex states\label{2.2}}
The corresponding wave function in configuration space is obtained
as follows. Using the relation
\begin{equation}\label{eq22}
\langle y\vert q\rangle = {e^{-y^2/2} H_q(y)\over \sqrt{2^q q!
\sqrt{\pi}}},\qquad H_q(y) = (-1)^q e^{y^2}
\partial_y^{(q)} e^{-y^2},\end{equation} and the
corresponding expression for $\langle x\vert N-q\rangle$, we
obtain
\begin{eqnarray}\label{eq23} \psi_{Nj}(x,y,t) & = & \langle x,y\vert U(t)\vert
N-j,j\rangle\nonumber\\& = & {e^{-(x^2+y^2)/2}\over \sqrt{\pi
2^N}}\sum_{q=0}^N C^{(q)}_{Nj}{H_{N-q}(x) H_q(y)
\over\sqrt{(N-q)!q!}}\nonumber\\& = & {(-1)^N e^{(x^2+y^2)/2}\over
\sqrt{\pi 2^N}}\sum_{q=0}^N C^{(q)}_{Nj}{\partial_x^{N-q}
\partial_y^q e^{-(x^2+y^2)}\over\sqrt{(N-q)!q!}
}\end{eqnarray} The wave function has a more appealing form in
polar coordinates as shown below. Writing $x=r\cos\theta$,
$y=r\sin\theta$, and defining
\begin{equation}\label{eq26}
\gamma_{\pm}(\tau)=v_{11}+v_{12}\tau \pm
i(v_{21}+v_{22}\tau),\end{equation} we get (see appendix B)
\begin{equation}\label{eq28} \psi_{Nj}(x,y,t)= \sum_{n=0}^N
b_{Nj}^{(n)} u_{N-n,n}(r,\theta)\end{equation} where
\begin{eqnarray}\label{eq29} b_{Nj}^{(n)} & = & {1\over j!} \sqrt{{(N-j)!
j!\over (N-n)! n! 2^N}}\zeta_{Nn}^{(j)}(0)\nonumber\\
\zeta_{Nn}(\tau)& = & \gamma_+(\tau)^{N-n}
\gamma_-(\tau)^n\nonumber\\ \zeta_{Nn}^{(j)}(0)& = &
\partial_\tau^j \zeta_{Nn}(\tau)\rfloor_{\tau\to 0}\nonumber\\
u_{mn}(r,\theta)& = & u^{LG}_{mn}(x,y,\sqrt{2}).\end{eqnarray}
Recall that for $m\neq n$, $u_{mn}(r,\theta)$ represents a vortex of
order $\vert m-n\vert$ and charge $m-n$ embedded in a Gaussian host
beam of waist $\omega=\sqrt{2}$. Thus {\it for odd values of $N$,
the wave function $\psi_{Nj}(x,y,t)$ becomes a coherent
superposition of vortex states, whereas for even values of $N$, the
superposition will also contain a state (corresponding to $n=N/2$)
 that does not have a vortex character.}\cite{zeilinger}.
\subsection{Creation of a single quantum vortex\label{2.3}}
In this section we will derive conditions for the creation of a
single quantum vortex. We reiterate that for light fields, the
vortex will appear in the quadrature distribution whereas for
other systems it will be in the probability distribution in
configuration space.

The initial two-mode Fock state can evolve into a single vortex
state when the summation in (\ref{eq28}) collapses into a single
term. This happens whenever $\gamma_+(0)$ or $\gamma_-(0)$ is
zero. It is easy to show that $\vert v_{21}\vert^2=1/2$ for both
these cases.

If $\gamma_+(0)=0$, then $v_{11}+i v_{21}=0$ and taking the
complex conjugate of this equation, $v_{22}+i v_{12}=0$. Then
$\gamma_+(\tau)=2 v_{12}\tau$ and $\gamma_-(\tau)=-2 i v_{21}$ so
that $\zeta_{Nn}^{(j)}(0)=(2 v_{12})^{N-n}(-2 i v_{21})^n j!
\delta_{N-j,n}$ and, finally \begin{equation}\label{eq30}
\psi_{Nj}(x,y,t)\rfloor_{\gamma_+(0)=0}=2^{N/2} i^{j-N}
v_{21}^{N-j} v_{12}^j u_{j,N-j} (r,\theta).\end{equation} The
condition $\gamma_+(0)=0$ implies that $\Omega=-g\sin \phi$ and
$\sigma \cos \sigma t=-g \cos\phi \sin\sigma t$. We give two
examples for which these conditions are
satisfied.\begin{enumerate}
\item Setting $\Omega=0$, $\phi=\pi$ and $\sigma t=\pi/4$, we get
\begin{equation}\label{eq31} \psi_{Nj}(x,y,t)=i^j u_{j,N-j}(r,\theta).\end{equation} From
(\ref{eq15}), one obtains the corresponding state vector
\begin{equation}\label{eq32} U_0\vert N-j,j\rangle=\sum_{q=0}^N D^{(q)}_{Nj}
\vert N-q,q\rangle\end{equation} where
\begin{equation}\label{eq33} U_0=\exp[{i \pi\over 4} (a^\dagger b
+ ab^\dagger)]\end{equation} and \begin{eqnarray}\label{eq34}
D^{(q)}_{Nj}& = & C^{(q)}_{Nj}\rfloor_{\Omega=0, \phi=\pi,\sigma t
=\pi/4}\nonumber\\ & = & \sqrt{{(N-j)! j!\over 2^N (N-q)! q!}}
{i^q\over j!}\left[
\partial_\tau^j (1+i\tau)^{N-q} (1-i\tau)^q\right]_{\tau\to 0}.\end{eqnarray}
\item Setting $\Omega=g$, $\phi=-\pi/2$ and $\sigma t=\pi/2$, we get
\begin{equation}\label{eq35} \psi_{Nj}(x,y,t)=(-i)^{j+N}
u_{j,N-j}(r,\theta).\end{equation} The operator form of the
corresponding state vector is given by \cite{simonnote}
\begin{equation}\label{eq36} \exp[{i \pi\over 2\sqrt{2}}
\{i(a^\dagger b - ab^\dagger)-(a^\dagger a-b^\dagger b)\}] \vert
N-j,j\rangle.\end{equation}
\end{enumerate}
 Following a similar
analysis for $\gamma_-(0)=0$, one obtains
\begin{equation}\label{eq37}
\psi_{Nj}(x,y,t)\rfloor_{\gamma_-(0)=0}=2^{N/2} i^{N-j}
v_{21}^{N-j} v_{12}^j u_{N-j,j} (r,\theta).\end{equation} The
condition $\gamma_-(0)=0$ yields $\Omega=g\sin \phi$ and $\sigma
\cos \sigma t=g \cos\phi \sin\sigma t$. These two conditions are
satisfied , for example, when $\Omega=\phi=0$ and $\sigma
t=\pi/4$. The corresponding wave function is the complex conjugate
of (\ref{eq31}).

We end this section by noting that the above conditions can be
physically realized for a given entangling device. We give an
example in the context of a frequency converter. Suppose the
signal (of frequency $\omega_a$) and the idler (of frequency
$\omega_b$) are initially in Fock states and the converter is
pumped at the difference frequency $\omega_a-\omega_b$. Replacing
$t$ by $L/c$, where $L$ is the length of the non-linear medium and
$c$ is the speed of light, one can adjust the pump amplitude such
that $gL/c=\pi/4$. This setup corresponds to $\Omega=\phi=0$ and
$gt=\pi/4$. In this case the quadrature distribution of the output
state will be a single quantum vortex as mentioned above.
\subsection{Entanglement of the two modes\label{2.4}}
Initially the two modes are not entangled as the state vector
$\vert N-j,j\rangle$ is the direct product of the state vectors
for each mode. In configuration space, this would imply that
$\psi_{Nj}(x,y,0)$ is separable in $x$ and $y$ as indeed it is.
Furthermore, as the time dependence arises solely in $v_{ij}$
which vary as $\cos \sigma t$ or $\sin \sigma t$, the initial
state is revived whenever $\sigma t=k\pi$ where $k$ is an integer.
For even values of $k$, the revival is exact whereas for odd
values of $k$, it is within an overall factor of $(-)^N$. At other
times, the two modes are entangled as is evident in the expression
((\ref{eq8}) or (\ref{eq15})) for the state vector and the
expression ((\ref{eq23}) or (\ref{eq28})) for the corresponding
wave function.
\subsection{Degree of entanglement\label{2.5}}
 Note
that the two-mode system $|\psi_{Nj}(t)\rangle$ is in a pure state
whereas the reduced state of each mode, determined by a partial
trace operation, will be a mixed state.  The reduced density
operators of modes `a' and `b' are given respectively by \numparts
\begin{eqnarray}\label{eq38} \rho^{(a)}_{Nj} & =Tr_b |\psi_{Nj}\rangle\langle
\psi_{Nj}|& =\sum_{q=0}^N |C^{(q)}_{Nj} |^2 |q\rangle\langle
q|\\\rho^{(b)}_{Nj} & =Tr_a |\psi_{Nj}\rangle\langle \psi_{Nj}|&
=\sum_{q=0}^N |C^{(N-q)}_{Nj}|^2 |q\rangle\langle q|
\end{eqnarray}
\endnumparts
The corresponding von Neumann entropies $S^{(a)}_{Nj}$ and
$S^{(b)}_{Nj}$ provide a measure of the degree of entanglement
between the two modes:
\numparts
\begin{eqnarray}\label{eq39}
S^{(a)}_{Nj}& = & -
\sum_{q=0}^N |C^{(q)}_{Nj} |^2 \log |C^{(q)}_{Nj} |^2\\
 \rule{0mm}{6mm}
S^{(b)}_{Nj}& = & - \sum_{q=0}^N |C^{(N-q)}_{Nj}|^2 \log
|C^{(N-q)}_{Nj}|^2 \end{eqnarray}
\endnumparts
By virtue of relations (\ref{eq20}), we get \be \label{eq40}
S^{(a)}_{Nj}\rfloor_{|v_{21}|^2\to 1-R}  =
S^{(a)}_{Nj}\rfloor_{|v_{21}|^2\to R} =
S^{(a)}_{N,N-j}\rfloor_{|v_{21}|^2\to R}.\ee
Changing the
summation index from $q$ to $N-q$ in the expression for
$S^{(b)}_{Nj}$, one obtains $S^{(a)}_{Nj}=S^{(b)}_{Nj}$. Thus the
symmetry relations (\ref{eq40}) hold good for $S^{(b)}_{Nj}$ as
well. These observations hold for any bipartite system in a pure
state.

It is remarkable that for a given value of $N$, $j$ and $q$, the
dynamics of $\vert C^{(q)}_{Nj}\vert^2$ depends on $\vert
v_{21}\vert^2=\sin^2\Theta\sin^2\sigma t$ only (see appendix A).
This important observation implies that {\it (a) the entropy
$S^{(a)}_{Nj}$ and the reduced density operator $\rho^{(a)}_{Nj}$
are independent of $\phi$ and (b) are symmetric with respect to
the interchange of $\Theta$ and $\sigma t$.} In Figure 1, we plot
$S^{(a)}_{Nj}$ as a function of $\vert v_{21}\vert^2$ for $N=4$
and $j=0, 1, 2$.
\begin{figure}
\setcaptionwidth{4.0in}
 \centering\includegraphics[width=4.0in]{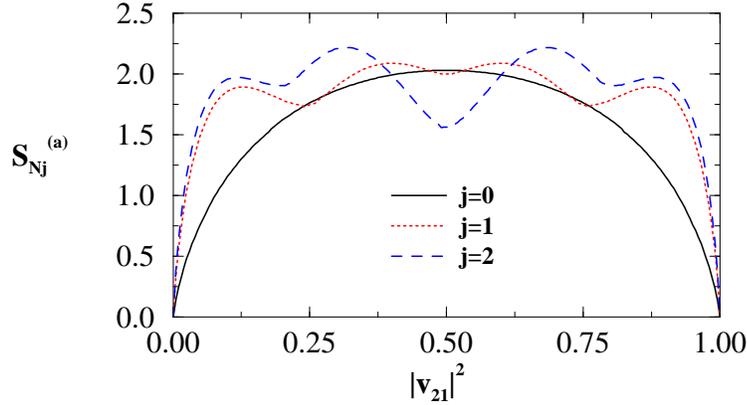}
\caption{\label{f1} Plot of $S^{(a)}_{Nj}$ as a function of $\vert
v_{21}\vert^2$ for $N=4$ and $j=0, 1, 2$.}
\end{figure}

Trivially, for $\vert v_{21}\vert^2=0$, the initial pure state
$|N-j,j\rangle$ either does not evolve or is fully revived and the
entropy of the reduced state is zero. For $\vert v_{21}\vert^2=1$,
the initial state swaps the photon numbers in the two modes and
becomes $|j,N-j\rangle$ which is also a pure state.  For all other
values of $\vert v_{21}\vert^2$, the initially pure state becomes
a mixed state and the entropy of the reduced state becomes
non-zero. Recall that for $\vert v_{21}\vert^2=1/2$ and $N-j\neq
j$, the quantum state becomes a vortex. Thus {\it a quantum vortex
is indeed an entangled state}. To quantify the degree of
entanglement for a vortex state, we plot $S^{(a)}_{Nj}$ as a
function of $j$ for $\vert v_{21}\vert^2=1/2$ and a given total
number of photons $N$ (see Fig. 2).
\begin{figure}
\setcaptionwidth{5.0in}
 \centering
\includegraphics[width=5.0in]{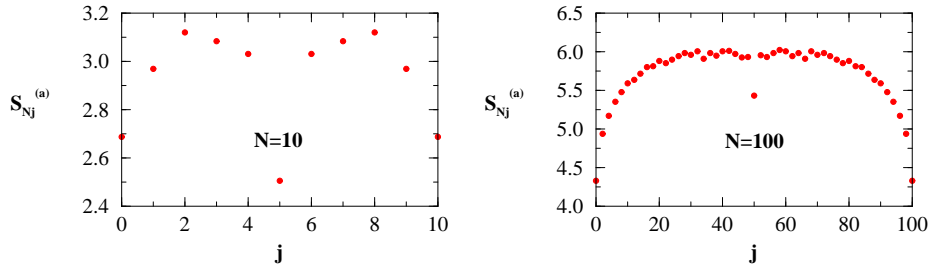}
\caption{\label{f2} Plot of $S^{(a)}_{Nj}$ as a function of $j$
for $\vert v_{21}\vert^2=1/2$ and a given total number of photons
$N$.}
\end{figure}
 It is clear that the entropy of the state without a vortex
($j=N/2$) is less than the entropy of the neighboring ($ j\sim N/2$)
vortex states ($N-j\neq j$). This reduction in entropy can be
attributed to the symmetry of the $j=N/2$ state and traced to the
highly oscillatory nature of the Jacobi polynomial appearing in Eq.
(\ref{eq21}). For a given value of $N$, the minimum in the entropy
of a {\it vortex state} occurs for $j=0,N$ in which case $\vert
C^{(q)}_{Nj}\vert^2$ is a binomial distribution (see appendix A).
Interestingly, for $j=0,N$, the vortex state will have the maximum
allowed order ($N$). Thus {\it the vortex state of maximum order
will have minimum entropy} which is counter-intuitive. One would
have expected that the more twists the phase of the state has, more
energetic and more entropic it would be. Note that the symmetry of
$S^{(a)}_{Nj}$ about $\vert v_{21}\vert^2=1/2$ in Fig. 1 and about
$j=N/2$ in Fig. 2 is contained in the relations (\ref{eq40}). Note
also that the vorticity or non-vorticity of the state of lowest
entropy depends on the value of $N$ (see Fig. 3).
\begin{figure}
\setcaptionwidth{4.0in}
 \centering\includegraphics[width=4.0in]{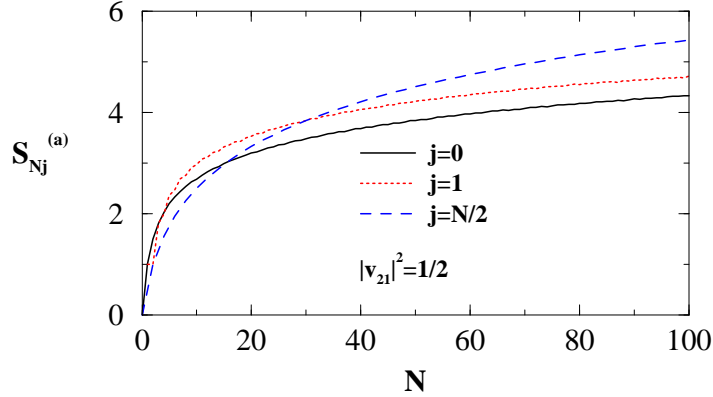}
\caption{\label{f3}Plot of $S^{(a)}_{Nj}$ as a function of $N$ for
$\vert v_{21}\vert^2=1/2$ and $j=0, 1, N/2$.}
\end{figure}

We end this section by comparing the entropy values in Figs. 1-3
with $\log_2 (N+1)$, the maximum entropy possible for a given $N$
with an entirely mixed state. For $N=4$, $10$ and $100$, $\log_2
(N+1)$ has the values $2.32193$, $3.45943$ and $6.65821$
respectively.
\section{Evolution of an initial vortex state\label{3}}
\subsection{Evolution of the state vector\label{3.1}}
Let us assume that the two modes are initially in a quantum vortex
state as in (\ref{eq32}). Then the state vector at time t will be
given by \begin{equation}\label{eq41} \vert \tilde{\psi}_{Nj}
(t)\rangle=U(t)U_0\vert N-j,j\rangle\end{equation} Proceeding as
in section \ref{2.1}, we define
\begin{equation}\label{eq42} {\hat{a}(t)\choose \hat{b}(t)}
=[U(t)U_0]^\dagger {\hat{a}\choose \hat{b}} [U(t)U_0]
\end{equation} and obtain \begin{equation}\label{eq43}
{\hat{a}(t)\choose \hat{b}(t)}= {\bf
V}{\hat{a}(0)\choose\hat{b}(0)}. \end{equation} Note, however,
that in this case, \begin{equation}\label{eq44} {\hat{a}(0)\choose
\hat{b}(0)} =U_0^\dagger {\hat{a}\choose \hat{b}}U_0= {\bf
W}{\hat{a}\choose\hat{b}}, \qquad {\bf W}={1\over
\sqrt{2}}\left(\begin{array}{cc}
  1 & i \\
  i & 1
\end{array}\right). \end{equation}
Thus \begin{equation}\label{eq46} {\hat{a}(t)\choose \hat{b}(t)}=
{\bf \tilde{V}}{\hat{a}\choose\hat{b}}; \quad {\bf \tilde{V}}=
{\bf V}{\bf W}.\end{equation} The action and the effect of the
unitary operator $U_0$ are now clear. $U_0$ transforms the
two-mode Fock state into a different initial state {\it before}
its time evolution begins and thus $U_0$ can be regarded as the
operator for {\it initial state preparation}. The effect of $U_0$
is contained in the unitary matrix ${\bf W}$. As a result, the
overall unitary evolution matrix changes from ${\bf V}$ to ${\bf
\tilde{V}}= {\bf V}{\bf W}$.

 In the present case, $U_0$, as given by (\ref{eq33}), prepares a quantum
vortex state as the initial state and the corresponding expression
for ${\bf W}$ is given as in (\ref{eq44}). A different expression
for $U_0$
 will generate an initial state that is different from a quantum
 vortex. Yet for all these initial states the dynamics is
 essentially solved once the corresponding dynamics for a two-mode
 Fock state is worked out as in section \ref{2.1}. In each case, one need only calculate
 the matrix ${\bf W}$ and replace ${\bf V}$ by ${\bf \tilde{V}}$.
{\it In this sense, our theory not only provides a unified
approach to entanglement through a generic Hamiltonian but also
promises wide applicability to a variety of initial states}.

In the present case, the matrix elements of ${\bf
\tilde{V}}=\{\tilde{v}_{ij}\}$ are obtained easily as
\begin{eqnarray}\label{eq47} \tilde{v}_{11} =
(v_{11}+iv_{12})/\sqrt{2}, &&\qquad \tilde{v}_{12} =  (v_{12}+i
v_{11})/\sqrt{2},\nonumber\\ \tilde{v}_{21} =  (v_{21}+i
v_{22})/\sqrt{2}, && \qquad \tilde{v}_{22}  =  (v_{22}+i
v_{21})/\sqrt{2}.\end{eqnarray} Using (\ref{eq7}), one can also
show that
\begin{equation}\label{eq48}\tilde{v}_{11}=\tilde{v}^*_{22},\quad
\tilde{v}_{12}=-\tilde{v}^*_{21}\quad\hbox{and}\quad \vert
\tilde{v}_{21}\vert^2+\vert \tilde{v}_{22}\vert^2=1.
\end{equation}

It is now trivial to obtain the wave vector and the wave function
by borrowing the corresponding results from the previous section.
We simply replace $v_{ij}$ by $\tilde{v}_{ij}$ for $i,j=1,2$ and
for the sake of clarity and comparison, use the same nomenclature
for the new expressions except for a $\tilde{ }$ (tilde) over
them. Thus \begin{equation}\label{eq49} \vert \tilde{\psi}_{Nj}
(t)\rangle=\sum_{q=0}^N \tilde{C}^{(q)}_{Nj} \vert N-q,q\rangle
\end{equation} where \numparts \begin{eqnarray}\label{eq50}
\tilde{C}^{(q)}_{Nj} & = &{1\over j!}\left[{(N-j)! j! \over
(N-q)!q!}\right]^{1/2}
\partial_\tau^{(j)} \tilde{\xi}_{N-q,q}(\tau)\rfloor_{\tau\to 0}\\
\tilde{\xi}_{pq}(\tau) & = & (\tilde{v}_{11}+\tilde{v}_{12}\tau)^p
(\tilde{v}_{21}+\tilde{v}_{22}\tau)^q\end{eqnarray}\endnumparts
Furthermore, the results of appendix A can be used to write
\begin{eqnarray}\label{eq52} \fl \vert \tilde{C}^{(q)}_{Nj}\vert^2 & =
 & (N-j)! (N-q)! q! (j!)^{-1} (1-|\tilde{v}_{21}|^2)^N
 \left({|\tilde{v}_{21}|^2\over 1-|\tilde{v}_{21}|^2}\right)^{q-j}\vert
 f^{(q)}_{Nj}(|\tilde{v}_{21}|^2)\vert^2\\\fl
&=& \label{eq53}\left\{\begin{array}{ll}
 \delta_{q,j}, & \mbox{$|\tilde{v}_{21}|^2\to 0$,}\\\rule{0mm}{6mm}
\delta_{q,N-j}, & \mbox{$|\tilde{v}_{21}|^2\to 1$.}
\end{array}
\right.\end{eqnarray} and \be \label{eq54}\vert
\tilde{C}^{(q)}_{Nj}\vert ^2\rfloor_{|\tilde{v}_{21}|^2\to 1-R}=
\vert \tilde{C}^{(N-q)}_{Nj}\vert
^2\rfloor_{|\tilde{v}\tilde{}_{21}|^2\to R} = \vert
\tilde{C}^{(q)}_{N,N-j}\vert ^2\rfloor_{|\tilde{v}_{21}|^2\to
R}.\ee
\subsection{The wave function\label{3.2}} The corresponding wave function in
configuration space can be read off from Eq (\ref{eq28}). We get
\begin{equation}\label{eq55} \tilde{\psi}_{Nj}(x,y,t)= \sum_{n=0}^N
\tilde{b}_{Nj}^{(n)} u_{N-n,n}(r,\theta)\end{equation} where
\numparts\begin{eqnarray}\label{eq56} \tilde{b}_{Nj}^{(n)} & = &
{1\over j!} \sqrt{{(N-j)! j!\over
(N-n)! n! 2^N}}\tilde{\zeta}_{Nn}^{(j)}(0)\\
\tilde{\zeta}_{Nn}(\tau)& = & \tilde{\gamma}_+(\tau)^{N-n}
\tilde{\gamma}_-(\tau)^n\\ \tilde{\gamma}_{\pm}(\tau) & = &
\tilde{v}_{11}+\tilde{v}_{12}\tau \pm
i(\tilde{v}_{21}+\tilde{v}_{22}\tau).\end{eqnarray}\endnumparts

Thus {\it a quantum vortex state evolves into a superposition of
vortex states} under the action of the Hamiltonian (\ref{eq1}).
\subsection{Revival and charge conjugation\label{3.3}}
It can be shown that if $\tilde{\gamma}_+(0)$ or
$\tilde{\gamma}_-(0)$ is zero, then the summation in (\ref{eq55})
reduces to a single term. Specifically, if
$\tilde{\gamma}_+(0)=0$, then ${\rm Im}\,v_{21}= {\rm Im
}\,v_{22}=0$ and \begin{equation}\label{eq57}
\tilde{\psi}_{Nj}(x,y,t)= (iv)^j {v^*}^{N-j}
u_{j,N-j}(r,\theta)\end{equation} with $v={\rm Re}\,v_{22}+i {\rm
Re }\,v_{21}$. The above conditions are satisfied for the
following cases: (a) $\sin \sigma t=0$ for arbitrary values of
$\Theta$ and $\phi$. This includes the initial state ($t=0$) and
the state upon revival ($\sigma t = \pi$). (b)
$\sin\Theta=\sin\phi=1$ for arbitrary time. In this case the
initial vortex state becomes an eigenstate of the corresponding
Hamiltonian.

On the other hand, if $\tilde{\gamma}_-(0)=0$, then ${\rm Re
}\,v_{21}= {\rm Re }\,v_{22}=0$ and \begin{equation}\label{eq58}
\tilde{\psi}_{Nj}(x,y,t)= (v)^j (-iv^*)^{N-j}
u_{N-j,j}(r,\theta)\end{equation} with $v={\rm Im }\,v_{22}+i {\rm
Im }\,v_{21}$. Note that
$u_{N-j,j}(r,\theta)=u^*_{j,N-j}(r,\theta)$ and thus
$\tilde{\gamma}_-(0)=0$ is the condition for {\it `charge
conjugation'} or {\it `helicity reversal'} of the initial vortex
state. This condition is fulfilled whenever $\sin\Theta\sin\phi=0$
and $\cos\sigma t=0$.
\subsection{Degree of Entanglement in the superposition state
(\ref{eq55})\label{3.4}}
 The reduced density operator of mode 'a' and the
corresponding von Neumann entropy are given respectively by
\numparts\begin{eqnarray}\label{eq59} \tilde{\rho}^{(a)}_{Nj} & =
& Tr_b |\tilde{\psi}_{Nj}\rangle\langle \tilde{\psi}_{Nj}|=
\sum_{q=0}^N |\tilde{C}^{(q)}_{Nj} |^2 |q\rangle\langle
q|\\\label{eq60}\tilde{S}^{(a)}_{Nj}& = & - \sum_{q=0}^N
|\tilde{C}^{(q)}_{Nj} |^2 \log |\tilde{C}^{(q)}_{Nj}
|^2\end{eqnarray}\endnumparts

 It is clear that
$\tilde{S}^{(a)}_{Nj}$ will depend on $\vert
\tilde{v}_{21}\vert^2$ in exactly the same way as $S^{(a)}_{Nj}$
does on $\vert v_{21}\vert^2$ except that the form of $\vert
\tilde{v}_{21}\vert^2$ as a function of $\Theta$, $\phi$ and
$\sigma t$ is quite different from $\vert v_{21}\vert^2$. Notably,
$\vert \tilde{v}_{21}\vert^2$ depends on $\phi$ while $\vert
v_{21}\vert^2$ does not. Explicitly, \begin{equation}\label{eq61}
\vert \tilde{v}_{21}\vert^2={1\over 2}-\sin\Theta \sin\sigma t
(\cos\phi \cos \sigma
 t-\sin\phi\cos\Theta\sin\sigma t).\end{equation}
\subsection{A case of constant entropy\label{3.5}}
Note that if $\Theta=\phi=\pi/2$ or $\Theta=0$, then $\vert
\tilde{v}_{21}\vert^2=1/2$ so that $ \vert
\tilde{C}^{(q)}_{Nj}\vert^2$ and the entropy
$\tilde{S}^{(a)}_{Nj}$ will not evolve with time. The underlying
reason is as follows.

Using the expressions (\ref{eq62}) for the SU(2) generators, we
obtain,
\begin{equation}\label{eq65} J_3\vert
N-q,q\rangle ={N-2 q\over 2}\vert
 N-q,q\rangle\end{equation}
 Noting that $U_0$, as given by \ref{eq33}), can be written as $U_0=\exp(i\pi J_1/2)$ and using the
 relation $J_2=U_0 J_3 U_0^\dagger$, we also get
 \begin{equation}\label{eq66}
 J_2U_0\vert
 N-j,j\rangle={N-2j\over 2}U_0\vert
 N-j,j\rangle\end{equation}
 For $\Theta=\phi=\pi/2$, the Hamiltonian reduces to $H=-2g J_2$
for which $U_0\vert N-j,j\rangle$ becomes an eigenstate by virtue
of (\ref{eq66}).

The condition $\Theta=0$ corresponds to $g=0$ and the Hamiltonian
 reduces to $2\Omega J_3$. From (\ref{eq32}), (\ref{eq41}) and (\ref{eq49}), one then immediately obtains
 $\tilde{C}^{(q)}_{Nj}=\exp(-i\Omega t[N-2q]) D^{(q)}_{Nj}$ so that $ \vert
\tilde{C}^{(q)}_{Nj}\vert^2$, and consequently, the entropy
$\tilde{S}^{(a)}_{Nj}$ become independent of time. The
corresponding
 wave function is given by (\ref{eq55}) where the coefficients $\tilde{b}^{(n)}_{Nj}$ have
 the value
 \be\label{eq67} \fl \tilde{b}^{(n)}_{Nj} = {N!\over j!} {N\choose j}^{-1/2} {N\choose n}^{-1/2}
 (-i)^{N-n} (\sin\Omega t)^{N-n+j} (\cos\Omega t)^{n-j} f^{(n)}_{Nj}(\cos^2\Omega
 t).\ee
 It is interesting that although $\vert \tilde{C}^{(q)}_{Nj}\vert^2$ and the
 entropy $\tilde{S}^{(a)}_{Nj}$ remain constant for $\Theta=0$,
 the initial
 vortex state will continue to evolve with time as shown in Figure
 4 \cite{oe}.
\begin{figure}
\setcaptionwidth{5.0in}
 \centering
 \includegraphics[width=5.0in]{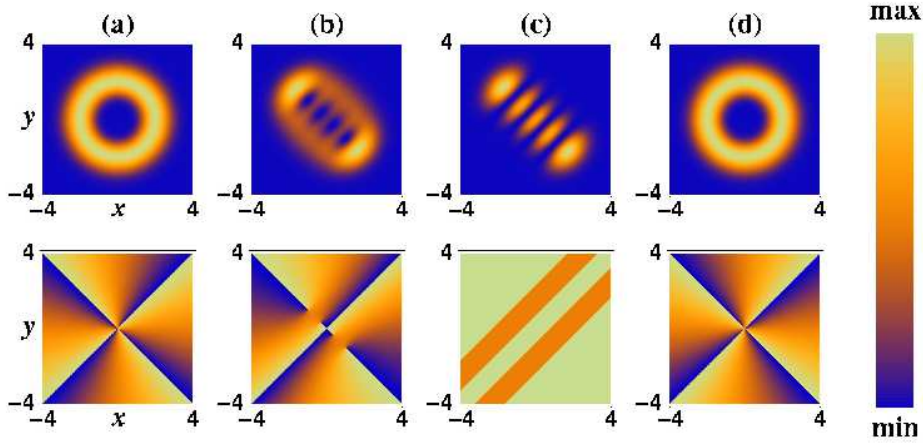}
\caption{\label{f4}Time evolution of an initial vortex state for
$\Theta=0$ even though the entropy $\tilde{S}^{(a)}_{Nj}$ remains
constant. Shown here are the contour plots of the absolute square
(top row) and the phase (bottom row) of $\tilde{\psi}_{Nj}(x,y,t)$
as functions of $x$ and $y$ at different times with $b^{(n)}_{Nj}$
as given in (\ref{eq67}). Here, $N=4$, $j=0$ and $\cos^2\Omega t$
has the values (a) 1.0, (b) 0.9, (c) 0.5 and (d) 0.0. The
horizontal and the vertical axes refer to the $x$ and $y$
coordinates respectively. For the phase plots, we have used the
convention that the phase ranges from $-\pi$ to $\pi$. Note that
for $\cos^2\Omega t=0$ the state becomes the complex conjugate of
the initial state as the direction of phase change is reversed.}
\end{figure}
For $N=4$ and $j=0$, the initial state at $t=0$ (Figure 4(a))
corresponding to $\cos^2\Omega t=1.0$ is a vortex of order 4 and
charge -4 as given by equations (\ref{eq31}) and (\ref{eq32}).
Recall that $\Omega=\sigma \cos\Theta$. Thus $\Theta=0$
corresponds to $\Omega=\sigma$. Furthermore, if $\cos^2\Omega
t=0.0$, then, with $\Theta=0$, the conditions for charge
conjugation as given below equation (\ref{eq58}) are satisfied and
we obtain the complex conjugate of the initial vortex. For
$\cos^2\Omega t=0.5$, it is more convenient to use cartesian
co-ordinates. Using the expression for $\tilde{C}^{(q)}_{Nj}$ as
given above and the configuration space representation of number
states as given by (\ref{eq22}), one can use the summation theorem
for Hermite polynomials \cite{gradshteyn} to obtain \numparts
\begin{eqnarray}\label{eq681}\fl \tilde{\psi}_{40}(x,y,t)\rfloor_{\Omega
t=\frac{\pi}{4}} & = & - {e^{-(x^2+y^2)/2}\over \sqrt{\pi 2^4 4!}}
H_4\left(\frac{x-y}{\sqrt{2}}\right)\\& = & {e^{-(x^2+y^2)/2}\over
\sqrt{ 24 \pi}} [ -(x-y)^4+6 (x-y)^2 -3]\end{eqnarray}
\endnumparts
Thus, the wave function is a Gaussian modulated by a Hermite
polynomial. Clearly, its value is real and, therefore, its phase
is either zero or $\pi$ depending respectively on whether the wave
function is $\geq 0$ or negative. It is easy to show that the wave
function vanishes whenever $(x-y)^2=3\pm \sqrt{6}$. Finally, it
may be of some interest to realize that the wave function
corresponds to a SU(2) coherent state $-\vert \tau, N\rangle$ in
the Schwinger representation \cite{oe} with $\tau=-1$ and $N=4$.
\section{Structure of the reduced state\label{4}}
 In order to determine the structure of the reduced state
for each mode, we first consider the correlation function in the
$x$-space of mode `a' given by $\langle x\vert
\rho^{(a)}_{Nj}\vert y\rangle$. A classical analog of this
function is the mutual coherence function of a partially coherent
source \cite{classvortex}. Thus the process of reduction of a pure
two-mode state into a mixed state by a partial trace operation
over one mode amounts to loss of coherence and information. We can
also define the spatial coherence function $\gamma^{(a)}_{Nj}(l)$
for the reduced state by \begin{equation}\label{eq68}
\gamma^{(a)}_{Nj}(l)= \int \langle x\vert \rho^{(a)}_{Nj}\vert
x+l\rangle \,dx\end{equation}

When the system is initially in a two-mode Fock state, one obtains
\begin{equation}\label{eq69} \langle x\vert \rho^{(a)}_{Nj}\vert
y\rangle=\sum_{q=0}^N {\vert C^{(q)}_{Nj}\vert^2 \over 2^q q!
\sqrt{\pi}} e^{-(x^2+y^2)/2} H_q(x) H_q(y).\end{equation} The
corresponding expression for $\gamma^{(a)}_{Nj}(l)$ is obtained by
evaluating the standard integral \cite{gradshteyn} in
(\ref{eq68}). We get
\begin{equation}\label{eq70} \gamma^{(a)}_{Nj}(l)= \sum_{q=0}^N \vert
C^{(q)}_{Nj}\vert^2 e^{-l^2/4} L_q(l^2/2)\end{equation} where
$L_q(x)=L_q^0(x)$ is a Laguerre polynomial. Note that only one
term survives in the summations over $q$ by virtue of (\ref{eq19})
whenever $\vert v_{21}\vert^2=0$ or $1$. Thus
\begin{equation}\label{eq71} \langle x\vert \rho^{(a)}_{Nj}\vert
y\rangle ={e^{-(x^2+y^2)/2}\over \sqrt{\pi}}
\left\{\begin{array}{ll}
 {H_j(x) H_j(y)\over 2^j j!}, & \mbox{$|v_{21}|^2\to 0$,}\\
 \rule{0mm}{6mm}
{H_{N-j}(x) H_{N-j}(y)\over 2^{N-j} (N-j)!}, &
\mbox{$|v_{21}|^2\to 1$.}
\end{array}
\right. \end{equation} and \begin{equation}\label{eq72}
\gamma^{(a)}_{Nj}(l)=e^{-l^2/4}\left\{\begin{array}{ll}
 L_j(l^2/2), & \mbox{$|v_{21}|^2\to 0$,}\\
 \rule{0mm}{6mm}
L_{N-j}(l^2/2), & \mbox{$|v_{21}|^2\to 1$.}
\end{array}
\right. \end{equation} Furthermore, one can use Eq. (\ref{eq20})
and the definition (\ref{eq38}) to get \numparts \begin{eqnarray}
\label{eq73} \langle x\vert \rho^{(a)}_{Nj}\vert
y\rangle\rfloor_{|v_{21}|^2\to 1-R} & = & \langle
x\vert\rho^{(a)}_{N,N-j}\vert y\rangle\rfloor_{|v_{21}|^2\to R}
\\\label{eq74} \gamma^{(a)}_{Nj}(l)\rfloor_{|v_{21}|^2\to 1-R} & = &
\gamma^{(a)}_{N,N-j}\rfloor_{|v_{21}|^2\to
R}\end{eqnarray}\endnumparts

When the system is initially in a vortex state, Eqs.
(\ref{eq69}-\ref{eq74}) are still valid provided that $|v_{21}|^2$
is replaced by $|\tilde{v}_{21}|^2$.
\begin{figure}
\setcaptionwidth{5.0in}
 \centering
 \includegraphics[width=5.0in]{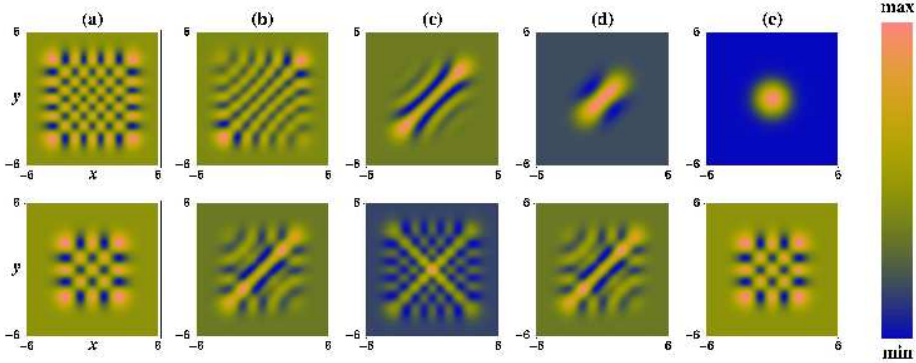}
\caption{\label{f5}Contour plots of the correlation function as a
function of $x$ and $y$ for different values of $|v_{21}|^2$. The
parameters are as follows: $N=8$; $j=0$ (top row), $j=4$ (bottom
row); $|v_{21}|^2$ has the values (a) $1.0$, (b) $0.9$, (c) $0.5$,
(d) $0.1$ and (e) $0.0$. The horizontal and the vertical axes
refer to the $x$ and $y$ coordinates respectively.}
\end{figure}
In Fig. 5 we present contour plots of the correlation function as
a function of $x$ and $y$ for a set of values of $|v_{21}|^2$
 when $N=8$ and $j=0$ (top
row), $j=4$ (bottom row). The intricate patterns for
$|v_{21}|^2=0$ and $1$ can be explained by using Eq. (\ref{eq71}).
Furthermore, the identical nature of patterns for $N=8$, $j=4$ on
either side of $|v_{21}|^2=1/2$ can be attributed to the property
(\ref{eq73}). Finally in Fig. 6 we plot $\gamma^{(a)}_{Nj}(l)$ as
a function of $l$ and $|v_{21}|^2$ when $N=4$ and $j=0$, $2$. The
patterns for $|v_{21}|^2=0$ and $1$ follow from Eq. (\ref{eq72})
and the symmetry of the plot for $j=2$ about $|v_{21}|^2=1/2$
follows from (\ref{eq74}).
\begin{figure}
\setcaptionwidth{5.0in}
 \centering
 \includegraphics[width=5.0in]{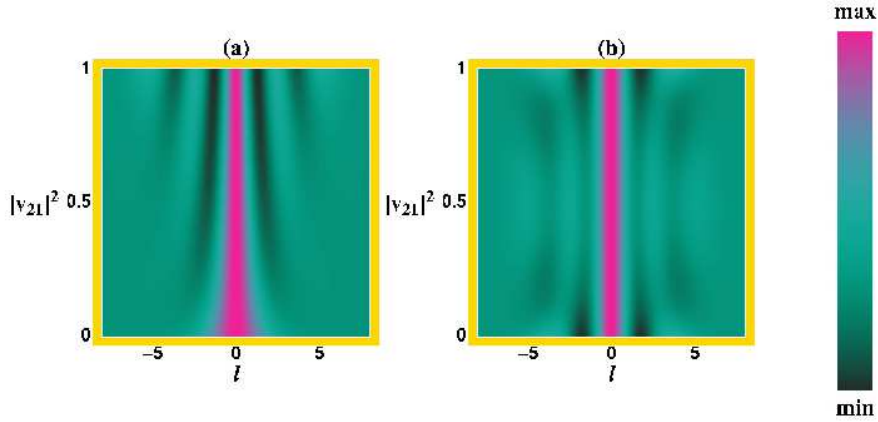}
\caption{\label{f6} Contour plot of $\gamma^{(a)}_{Nj}(l)$ as a
function of $l$ and $|v_{21}|^2$ when $N=4$ and $j=0$ (left),
$j=2$ (right).}
\end{figure}
\section{Conclusion\label{5}}
In conclusion, we have studied, in a general way, entanglement
produced in a two-mode bosonic system by linear SU(2)
transformations leading to the generation and evolution of quantum
vortex states. The linear SU(2) transformations are generated by
evolving the system under the action of a generic Hamiltonian that
mimics a variety of entanglement mechanisms. We have demonstrated
that these transformations produce a coherent superposition of
quantum vortices in general, and a single quantum vortex under
certain conditions. Furthermore, as one would expect, a vortex state
is found to be an entagled state. When the system is a light field,
the vortex will appear in the quadrature distribution that can be
measured by a homodyne method \cite{leonhardt}. Explicit analytical
results were obtained when the system was initially either in a Fock
state or in a quantum vortex state. In the latter case, we have also
found conditions for its revival and charge conjugation. A simple
recipe was provided to accommodate all other cases for which the
initial state can be reached from a Fock state by a unitary
transformation. Thus we not only provide a unified approach to
entanglement through a generic Hamiltonian but also predict wide
applicability of our results to a variety of initial states.

The ideas developed in this paper can be applied not only to light
fields but also to matter waves such as the Bose Einstein
condensates (BEC). In recent years, the BEC has proved to be an
excellent laboratory for studying (both bipartite and many-particle)
entanglement \cite{bec1,bec2,bec3,deb2,deb3}. The entanglement of
the modes as well as the entanglement of the atoms in a BEC have
been considered. We mention parenthetically that our work is
relevant in the former case.  It is also well known that several
mechanisms exist for the generation of vortices in a BEC
\cite{becv,schleich1,schleich2,kapale}. Additionally, Whyte et al.
\cite{whyte} have used the similarity between BECs and laser light
to propose a method for generating Hermite-Gaussian type modes in a
so-called {\it light pulse resonator}. Thus it should indeed be
possible to generate vortices in a two-component BEC by entangling
the two modes of the BEC by linear SU(2) transformations via an
entangling device such as a beam splitter.
\ack
One of us (GSA) would like to thank NSF for supporting this
work under grant no. CCF-0524673.
\appendix
\section{Some useful properties of $\vert C^{(q)}_{Nj}\vert^2$ }
 Using Leibniz' rule for the $j$-th derivative of
a product and the relations (\ref{eq7}), one obtains
\be\label{eq17} \fl \vert C^{(q)}_{Nj}\vert^2 =
 (N-j)! (N-q)! q! (j!)^{-1} (1-|v_{21}|^2)^N
\left({|v_{21}|^2\over 1-|v_{21}|^2}\right)^{q-j}\vert
f^{(q)}_{Nj}(|v_{21}|^2)\vert^2\ee with
\begin{equation}\label{eq18} f^{(q)}_{Nj}(|v_{21}|^2)=\sum_{k=0}^j {(-1)^k
{j\choose k} \over (N-q-k)! (q-j+k)!}\left({|v_{21}|^2\over
1-|v_{21}|^2}\right)^k .\end{equation}  It is easy to show that
\begin{equation}\label{eq19} \vert C^{(q)}_{Nj}\vert^2 =
\left\{\begin{array}{ll}
 \delta_{q,j}, & \mbox{$|v_{21}|^2\to 0$,}\\ \rule{0mm}{6mm}
\delta_{q,N-j}, & \mbox{$|v_{21}|^2\to 1$.}
\end{array}
\right. \end{equation} Next we derive some important symmetry
properties of $\vert C^{(q)}_{Nj}\vert^2$. First we show that
\begin{eqnarray*} f^{(q)}_{Nj}(1-R) & = & (-1)^j\left({1-R\over
R}\right)^j f^{(N-q)}_{Nj}(R)\\ f^{(q)}_{N,N-j}(R) & = &
(-1)^{N-q}{(N-j)!\over j!}\left({R\over
1-R}\right)^{N-q}f^{(q)}_{Nj}(1-R)\end{eqnarray*} where $0\leq
R\leq 1$. The first relation is obtained from (\ref{eq18}) by
changing the summation index from $k$ to $j-k$ and the second
relation is proved by exploiting the non-negativity of the
factorials in (\ref{eq18}) and changing the summation range
accordingly. Using these two relations we immediately get \be
\label{eq20}\vert C^{(q)}_{Nj}\vert ^2\rfloor_{|v_{21}|^2\to 1-R}
 =  \vert C^{(N-q)}_{Nj}\vert ^2\rfloor_{|v_{21}|^2\to
R}= \vert C^{(q)}_{N,N-j}\vert ^2\rfloor_{|v_{21}|^2\to R}.\ee

 Note that if $|v_{21}|^2=1/2$, then
$|v_{11}|^2=|v_{22}|^2=|v_{12}|^2=1/2$ as well. Additionally, if
$j=0$ or $N$, then  $\vert C^{(q)}_{Nj}\vert^2=2^{-N} {N\choose
q}$ is a binomial distribution whereas if $j=N/2$, then
 \begin{equation}\label{eq21} \vert
 C^{(q)}_{N/2,N/2}\vert^2 = (N!)^{-1}[(N/2)! P_{N/2}^{(N/2
-q,q-N/2)}(0)]^2 {N\choose q} \end{equation} where
$P_n^{(\alpha,\beta)}(x)$ is a Jacobi polynomial.
\section{Derivation of equation (\ref{eq28})}
Substituting the expression (\ref{eq16}) for $C^{(q)}_{Nj}$ in
(\ref{eq23}) and performing the summation over $q$ {\it before}
differentiation with respect to $\tau$, we get \be\label{eq24} \fl
\psi_{Nj}(x,y,t)= {(-1)^N\over N!} \sqrt{{(N-j)!\over j! 2^N
\pi}}e^{(x^2+y^2)/2} [\partial_\tau^{(j)}
\hat{A}^N(\tau)e^{-(x^2+y^2)}]_{\tau\to 0}\ee where
\begin{equation}\label{eq25}\hat{A}(\tau)=(v_{11}
+v_{12}\tau)\partial_x+(v_{21}
+v_{22}\tau)\partial_y.\end{equation} We introduce $z=x+iy$ so
that $x^2+y^2=zz^*$ and
$\hat{A}(\tau)=\gamma_{+}(\tau)\partial_z+\gamma_{-}(\tau)\partial_{z^*}$
with $\gamma_{\pm}$ given by (\ref{eq26}). Next we expand
$\hat{A}(\tau)^N$ binomially and then use the relation
\begin{equation}\label{eq27}
\partial^m_z\partial^n_{z^*} e^{-zz^*}=\left\{\begin{array}{ll}
 (-1)^n m! e^{-zz^*} z^{n-m}L_m^{n-m}(zz^*), & \mbox{$m\leq
 n$,}\\
 \rule{0mm}{6mm}
(-1)^m n! e^{-zz^*} {z^*}^{m-n}L_n^{m-n}(zz^*), & \mbox{$n\leq
m$.}
\end{array}
\right. \end{equation} to evaluate $\hat{A}(\tau)^N
e^{-(x^2+y^2)}$. Collecting all the terms  we finally obtain
(\ref{eq28}).


\begin{thebibliography}{99}
\bibitem{simonnote}Although relation (\ref{eq32}) has been derived earlier in a different
context \cite{simon}, relation (\ref{eq36}) has not been reported
in the literature to the best of our knowledge.
\end{thebibliography}
\end{document}